\documentclass[conference]{IEEEtran}
\IEEEoverridecommandlockouts
\usepackage[T1]{fontenc} %
\usepackage{amsmath}
\usepackage[cmintegrals]{newtxmath}
\usepackage{bm} %
\usepackage{url}
\usepackage{mdwtab}
\usepackage{siunitx}
\usepackage{xcolor}
\makeatletter
\let\MYcaption\@makecaption
\makeatother
\usepackage[font=footnotesize]{subcaption}

\makeatletter
\let\@makecaption\MYcaption
\makeatother

\DeclareMathOperator{\E}{\mathbb{E}}
\usepackage[acronym,nopostdot,style=index,nonumberlist,toc]{glossaries}
\usepackage{graphicx}

\usepackage{svg}
\usepackage{mathtools}
\usepackage{makecell}

\usepackage{breqn}   %

\usepackage{booktabs}
\usepackage{tabularx}
\newacronym{mmwave}{mmWave}{millimeter wave}
\newacronym{ai}{AI}{artificial intelligence}
\newacronym{embb}{eMBB}{enhanced mobile broadband}
\newacronym{urllc}{uRLLC}{ultra-reliable and low-latency communications}
\newacronym{mmtc}{mMTC}{massive machine-type communications}
\newacronym{nr}{NR}{new radio}
\newacronym{5g}{5G}{fifth generation}
\newacronym{3gpp}{3GPP}{Third Generation Partnership Project}
\newacronym{bs}{BS}{base station}
\newacronym{ue}{UE}{user equipment}
\newacronym{dci}{DCI}{downlink control information}
\newacronym{uci}{UCI}{uplink control information}
\newacronym{dcm}{DCM}{downlink control message}
\newacronym{ucm}{UCM}{uplink control message}
\newacronym{pdcch}{PDCCH}{physical downlink control channel}
\newacronym{pdsch}{PDSCH}{physical downlink shared channel}
\newacronym{pucch}{PUCCH}{physical uplink control channel}
\newacronym{pusch}{PUSCH}{physical uplink shared channel}
\newacronym{rrm}{RRM}{radio resource management}
\newacronym{dlsch}{DL-SCH}{downlink shared channel}
\newacronym{bch}{BCH}{broadcast channel}
\newacronym{pch}{PCH}{paging channel}
\newacronym{ul}{UL}{uplink}
\newacronym{kpi}{KPI}{key performance indicator}
\newacronym{dl}{DL}{downlink}
\newacronym{ulsch}{UL-SCH}{uplink shared channel}
\newacronym{rach}{RACH}{random-access channel}
\newacronym{mac}{MAC}{medium access control}
\newacronym{tdma}{TDMA}{time division multiple access}
\newacronym{tti}{TTI}{transmission time interval}
\newacronym{phy}{PHY}{physical layer}
\newacronym{bler}{BLER}{block error rate}
\newacronym{tb}{TB}{transport block}
\newacronym{tbs}{TBS}{transport block size}
\newacronym{tbler}{TBLER}{transport block error rate}
\newacronym{prb}{PRB}{physical resource block}
\newacronym{re}{RE}{resource element}
\newacronym{rb}{RB}{resource block}
\newacronym{sr}{SR}{scheduling request}
\newacronym{sg}{SG}{scheduling grant}
\newacronym{ack}{ACK}{acknowledgement}
\newacronym{rl}{RL}{reinforcement learning}
\newacronym{ml}{ML}{machine learning}
\newacronym{mdp}{MDP}{Markov decision process}
\newacronym{ql-amc}{QL-AMC}{Q-learning based adaptive modulation and coding}
\newacronym{ql-la}{QL-LA}{Q-learning based link adaptation}
\newacronym{dqn}{DQN}{deep Q-network}
\newacronym{ma}{MA}{multi-agent}
\newacronym{marl}{MARL}{multi-agent reinforcement learning}
\newacronym{mas}{MAS}{multi-agent systems}
\newacronym{maddpg}{MADDPG}{multi-agent deep deterministic policy gradient}
\newacronym{matd3}{MATD3}{multi-agent twin delayed deep deterministic policy gradient}
\newacronym{ddpg}{DDPG}{deep deterministic policy gradient}
\newacronym{ctde}{CTDE}{centralized training and decentralized execution}
\newacronym{sdu}{SDU}{service data unit}
\newacronym{pdu}{PDU}{protocol data unit}
\newacronym{ci}{CI}{confidence interval}
\newacronym{mlp}{MLP}{multilayer perceptron}
\newacronym{relu}{ReLU}{rectified linear unit}
\newacronym{dec-pomdp}{Dec-POMDP}{decentralized partially observable Markov decision process}
\newacronym{sc}{SC}{speaker consistency}
\newacronym{ic}{IC}{instantaneous coordination}
\newglossary[nlg]{notation}{not}{ntn}{Notation}

\newglossaryentry{not:nUE}{
   name=\ensuremath{L},
   description={number of users},
   type=notation
   }

\newglossaryentry{not:buffer-size}{
   name=\ensuremath{B},
   description={buffer size},
   type=notation
   }

\newglossaryentry{not:bitlength}{
   name=\ensuremath{\Upsilon},
   description={bitlength},
   type=notation
   }

\newglossaryentry{not:policy-net}{
   name=\ensuremath{\mu},
   description={policy network},
   type=notation
   }

\newglossaryentry{not:mutual-information}{
   name=\ensuremath{I},
   description={mutual information},
   type=notation
   }

\newglossaryentry{not:policy-params}{
   name=\ensuremath{\theta},
   description={policy network},
   type=notation
   }

\newglossaryentry{not:avg-packets}{
   name=\ensuremath{\lambda},
   description={average number of packets to transmit},
   type=notation
   }

\newglossaryentry{not:reward-signal}{
   name=\ensuremath{\rho},
   description={number of packets to transmit},
   type=notation
   }

\newglossaryentry{not:p-arrival}{
   name=\ensuremath{p_{\mathrm{a}}},
   description={probability of arriving a new packet},
   type=notation
   }

\newglossaryentry{not:episode-duration}{
   name=\ensuremath{T},
   description={maximum number of TTIs},
   type=notation
   }

\newglossaryentry{not:time-step}{
   name=\ensuremath{t},
   description={time step},
   type=notation
   }

\newglossaryentry{not:dl-vocabulary-size}{
   name=\ensuremath{D},
   description={downlink control message vocabulary size},
   type=notation
   }

\newglossaryentry{not:ul-vocabulary-size}{
   name=\ensuremath{U},
   description={uplink control message vocabulary size},
   type=notation
   }

\newglossaryentry{not:dl-message}{
   name=\ensuremath{m},
   description={downlink control message},
   type=notation
   }

\newglossaryentry{not:ul-message}{
   name=\ensuremath{n},
   description={uplink control message},
   type=notation
   }

\newglossaryentry{not:ue-action}{
   name=\ensuremath{a},
   description={UE action},
   type=notation
   }

\newglossaryentry{not:obs}{
  name=\ensuremath{o},
  description={observation},
  type=notation
  }

\newglossaryentry{not:state}{
  name=\ensuremath{x},
  description={agent state},
  type=notation
  }

\newglossaryentry{not:history-len}{
  name=\ensuremath{K},
  description={length of history buffer},
  type=notation
  }

\newglossaryentry{not:episodes-train}{
  name=\ensuremath{N_\mathrm{train}},
  description={number of training episodes},
  type=notation
  }

\newglossaryentry{not:episodes-eval}{
  name=\ensuremath{N_\mathrm{eval}},
  description={number of evaluation episodes},
  type=notation
  }

\newglossaryentry{not:episodes-test}{
  name=\ensuremath{N_\mathrm{test}},
  description={number of test episodes},
  type=notation
  }

\newglossaryentry{not:repetitions}{
  name=\ensuremath{N_\mathrm{rep}},
  description={number of test episodes},
  type=notation
  }

\newglossaryentry{not:n-receptions}{
  name=\ensuremath{N_\mathrm{SDUs}},
  description={number of test episodes},
  type=notation
  }

\newglossaryentry{not:n-collisions}{
  name=\ensuremath{N_\mathrm{c}},
  description={number of collisions},
  type=notation
  }

\renewcommand\baselinestretch{.98}

\newcommand{\nonl}{\renewcommand{\nl}{\let\nl\oldnl}}
\let\oldnl\nl%

\begin{document}

\title{Scalable Joint Learning of Wireless Multiple-Access Policies and their Signaling
}

\author{
  \IEEEauthorblockN{
    Mateus P. Mota\IEEEauthorrefmark{1}\IEEEauthorrefmark{2},
    Alvaro Valcarce\IEEEauthorrefmark{1},
    Jean-Marie Gorce\IEEEauthorrefmark{2},
    }
  \IEEEauthorblockA{
    \IEEEauthorrefmark{1}Nokia Bell Labs, Nozay, France\\
    Email: mateus.pontes_mota@nokia.com, alvaro.valcarce_rial@nokia-bell-labs.com
   }
    \IEEEauthorblockA{
    \IEEEauthorrefmark{2}National Institute of Applied Sciences, Lyon, France\\
    Email: jean-marie.gorce@insa-lyon.fr
    }
}

\maketitle

\begin{abstract}

    In this paper, we apply an \gls{marl} framework allowing the \gls{bs} and the \glspl{ue} to jointly learn a channel access policy and its signaling in a wireless multiple access scenario.
    In this framework, the \gls{bs} and \glspl{ue} are \gls{rl} agents that need to cooperate in order to deliver data.
    The comparison with a contention-free and a contention-based baselines shows that our framework achieves a superior performance in terms of goodput even in high traffic situations while maintaining a low collision rate.
    The scalability of the proposed method is studied, since it is a major problem in \gls{marl} and this paper provides the first results in order to address it.

\end{abstract}


\glsresetall

\section{Introduction}
\label{sec:Intro}

The goal of this paper is to explore a framework for jointly learning a channel access policy and its signaling policy for \gls{mac} in multiple-access scenarios.
This study aims at proposing a general framework capable of producing application-tailored protocols which may lead to performance gains over more general purpose protocols.

It is expected that \Gls{ai} and \gls{ml} will play a crucial role in 6G \cite{6g_roadmap} in making the network more adaptable and self-upgradable, helping meeting the requirements while also making the network management and optimization simpler.
One promising area in \gls{ml} for achieving a more adaptable network system is \gls{rl}.
In particular, \gls{marl} has been used to emerge communication that allows a better cooperative behavior \cite{foerster2016learning, sukhbaatar2016learning}.
The framework used in this paper leverages \gls{marl} to allow the network nodes to learn the channel-access policy and the communication needed to best collaborate with one another, thus also learning the signaling.

\paragraph*{Related Work}
\Gls{rl} has been used to develop channel access policies for the \gls{mac} in \cite{dutta2021Synthesis} and \cite{guo2022Access}.
It has also been used to select which \gls{mac} protocol to use \cite{gomes2020automatic} or which blocks to use \cite{pasandi2021}.
Differently from such works we propose to learn a channel access policy and its signaling.
The idea of learning a given protocol and its signaling has already been addressed in a previous work \cite{alvaro2021}, while in \cite{mota2021emergence} we proposed the framework for emerging a \gls{mac} protocol in a multiple access scenario.

\paragraph*{Contribution}
This paper extends the previous one \cite{mota2021emergence} in two ways:
\begin{enumerate}
    \item Traffic model: By using a Poisson process, instead of limiting the total number of \glspl{sdu}, making the new model more realistic. The Poisson process is used, for example, to model message arrivals in a packet data networks or the arrival of new telephone calls.
    \item Scalability study: By evaluating the scalability both in terms of arrival rate as well as \glspl{ue}.
\end{enumerate}
Since we propose to fully emerge a protocol for the \gls{bs} and \glspl{ue}, scalability may be an issue because the \gls{bs} needs to communicate with all \glspl{ue}.

This work is structured as follows.
Section~\ref{sec:system-model} describes the system model used and in Section~\ref{sec:proposed}, we present a new framework allowing the emergence of \gls{mac} protocols with \gls{marl}.
Finally, Section~\ref{sec:simulation} illustrates the performance of our algorithm with numerical results, where we compare the proposed solution with a baseline.
The main conclusions are drawn in Section~\ref{sec:conclusion}.

\section{System Model}
\label{sec:system-model}

Consider a single cell with a \gls{bs} serving \gls{not:nUE} \glspl{ue} operating according to a \gls{tdma} scheme, where each \gls{ue} needs to deliver data to the \gls{bs}.
Each \gls{ue} has a transmission buffer of capacity \gls{not:buffer-size} \gls{mac} \glspl{sdu} initially empty.
The \gls{sdu} arrival is modeled as a Poisson process with probability of arrival \gls{not:p-arrival}.
So, a new \gls{sdu} is added to the buffer with probability \gls{not:p-arrival}, until a maximum number \gls{not:episode-duration} of steps is achieved.
The average number of \glspl{sdu} arriving at each \gls{ue}'s buffer in any given episode of duration \gls{not:episode-duration} is then:
\begin{equation}
    \gls{not:avg-packets} = \gls{not:p-arrival} \gls{not:episode-duration}
\end{equation}
The network nodes can exchange information, using messages through the control channels.
In the remainder of this paper, we refer to the \gls{ue} \gls{mac} agent and the \gls{bs} \gls{mac} agent as \gls{ue} and \gls{bs} , respectively.

The channel for the uplink data transmission is modeled as a packet erasure channel, where a \gls{tb} is incorrectly received with a probability referred to as \gls{tbler}.
The \glspl{ue} use the same frequency resources on the \gls{ulsch}, which leads to possible collisions.
The \glspl{dcm} and \glspl{ucm} are transmitted over the \gls{dl} and \gls{ul} control channels, which are assumed to be dedicated and error free, so without any contention or collision.

We assume that the sets of possible \gls{dl} and \gls{ul} control messages have cardinality \gls{not:dl-vocabulary-size} and \gls{not:ul-vocabulary-size}, respectively.
For example, the \glspl{dcm} in an \gls{dl} control vocabulary of size $\gls{not:dl-vocabulary-size} = 4$ would have a bitlength $\gls{not:bitlength}_{\mathrm{DL}}$ of $\log_2 {\gls{not:dl-vocabulary-size}}=2$.

At each time step \gls{not:time-step}, the \gls{bs} can send one control message to each \gls{ue} and each \gls{ue} can send one control message to the \gls{bs} while being able to send data \glspl{pdu} through the \gls{ulsch}.
Furthermore, the \glspl{ue} can also delete a \gls{sdu} from the buffer at each time step.

We define the cellwide goodput $G$ (in \glspl{sdu}/TTIs) as the number of \gls{mac} \glspl{sdu} received by the \gls{bs} per unit of time. \Glspl{sdu} received by the \gls{bs} several times are only counted once:
\begin{equation}
    G = \frac{N_{\mathrm{RX}}} {\gls{not:episode-duration}}
\end{equation}
\noindent where $N_{\mathrm{RX}}$ represents the number of unique \glspl{sdu} received.
Since the \gls{bs} can only receive at most one \gls{sdu} per time step $N_{\mathrm{RX}} \leq 1$, the maximum cellwide goodput on average can be calculated as:
\begin{equation}\label{eq.:max-goodput}
    G_{\mathrm{max}} = \min{(\gls{not:p-arrival} \gls{not:nUE}, 1)}
\end{equation}
The collision rate $\Gamma$ is the number of steps in which a collision happened divided by the total number of time steps:
\begin{equation}
    \label{eq:delivery-rate}
    \Gamma = \frac{\gls{not:n-collisions}}{\gls{not:episode-duration}} \text{.}
\end{equation}
\noindent where $\gls{not:n-collisions}$ represents the total number of time steps in which at least two \glspl{sdu} collided.

\section{Emerging a MAC Protocol with MARL}
\label{sec:proposed}

\subsection{MARL Formulation}

We formulate the problem defined above as a \gls{marl} cooperative task, where the \gls{mac} layers of the network nodes (\glspl{ue} and \gls{bs}) are \gls{rl} agents that need to learn how to communicate with each other to solve an uplink transmission task.
In addition, the \gls{ue} agents need to learn when to send data through the \gls{ulsch} and when to delete an \gls{sdu}, in other words, to learn how to correctly manage the buffer.
In order to decide how to act, an agent needs to consider the messages received from the other agents. %
In addition, the \glspl{ue} also take into account their buffer status when taking actions, while the \gls{bs} takes into account the state of the \gls{ulsch}, i.e idle, busy or collision-free reception.

We model this problem as a \gls{dec-pomdp} \cite{oliehoek2008optimal}, augmented with communication.
A \gls{dec-pomdp} for $n$ agents is defined by the global state space $\mathcal{S}$, an action space $\mathcal{A}_1 , \ldots , \mathcal{A}_n $, and an observation space  $\mathcal{O}_1 , \ldots , \mathcal{O}_n $ for each agent.
In \gls{dec-pomdp}, an agent observation does not fully describe the environment state.
All agents share the same reward and the action space of each agent is subdivided into one environment action space and a communication action space.
The communication action represents the message sent by an agent and it does not affect the environment directly, but it may be passed to other agents.
In this work, the agent state $x_i$ may comprise not only the agent's current observation, but also previous observations, actions and received messages.

We use the following notations:
\begin{itemize}

    \item $\gls{not:obs}^{\mathrm{u}}_{t}$: Observation received by the $u$\textsuperscript{th} \gls{ue} at time step \gls{not:time-step}.
    \item $\gls{not:obs}^{\mathrm{b}}_{t}$: Observation received by the \gls{bs} at time step \gls{not:time-step}.
    \item $\gls{not:ul-message}^{\mathrm{u}}_{t}$: The \gls{ucm} sent from the $u^{\mathrm{th}}$ \gls{ue} at time step $t$.
    \item $\gls{not:dl-message}^{\mathrm{u}}_{t}$: The \gls{dcm} sent to the $u^{\mathrm{th}}$ \gls{ue} at time step $t$.
    \item $\gls{not:ue-action}^{\mathrm{u}}_{t}$: Environment action of the $u^{\mathrm{th}}$ \gls{ue} at time step $t$.
    \item $\gls{not:state}^{\mathrm{u}}_{t}$: Agent state of the $u^{\mathrm{th}}$ \gls{ue} at time step $t$.
    \item $\gls{not:state}^{\mathrm{b}}_{t}$: Agent state of the \gls{bs} at time step $t$.

\end{itemize}

\paragraph*{Observations}
The observation $\gls{not:obs}^{\mathrm{u}}_{t} \in \left\{ 0, \dots, \gls{not:buffer-size} \right\} $ is a integer representing the number of \glspl{sdu} in the buffer of the \gls{ue} $u$ at that time $t$.
Similarly, the observation $\gls{not:obs}^{\mathrm{b}}_{t} $ received by the \gls{bs} is a discrete variable with $\gls{not:nUE} + 2$ possible states:
\begin{equation}\label{eq.:base-obs}
    \gls{not:obs}^{\mathrm{b}}_{t} = \begin{cases}
    0, \text{ if the \gls{ulsch} is idle} \\
    \mathrm{u}, \text{ } \parbox[t]{.33\textwidth}{ if the \gls{ulsch} is detected busy with a single \gls{pdu} from \gls{ue} $\mathrm{u}$, correctly decoded} \\
    \gls{not:nUE} + 1, \text{ non-decodable energy in the \gls{ulsch}} \\
    \end{cases}
\end{equation}
\noindent where $\mathrm{u} \in \left\{ 0, \dots, \gls{not:nUE} \right\}$.

\paragraph*{Actions}
The environment action $\gls{not:ue-action}^{\mathrm{u}}_{t}  \in \{0, 1, 2\}$ is interpreted as follows:
\begin{equation}
\label{eq.:action}
    \gls{not:ue-action}^{\mathrm{u}}_{t} = \begin{cases}
    0 \text{: do nothing} \\
    1 \text{: transmit the oldest \gls{sdu} in the buffer} \\
    2 \text{: delete the oldest \gls{sdu} in the buffer} \\
\end{cases}
\end{equation}
We highlight that the \gls{dcm} and \gls{ucm} messages, \gls{not:dl-message} and \gls{not:ul-message}, are communication actions that the agents select while also being information available to the other agent's state as received message.

The agent state at time step $t$ is a tuple comprising the most recent $k$ observations, actions and received messages:
\begin{itemize}
    \item \Gls{ue} $\mathrm{u}$: $\gls{not:state}^{\mathrm{u}}_{t} = ( \gls{not:obs}^{\mathrm{u}}_{t}, \ldots , \gls{not:obs}^{\mathrm{u}}_{t-k} , \gls{not:ue-action}^{\mathrm{u}}_{t}, \ldots, \gls{not:ue-action}^{\mathrm{u}}_{t-k}, \gls{not:ul-message}^{\mathrm{u}}_{t}, \ldots, \gls{not:ul-message}^{\mathrm{u}}_{t-k} , \\ \gls{not:dl-message}^{\mathrm{u}}_{t}, \ldots, \gls{not:dl-message}^{\mathrm{u}}_{t-k} ) $
    \item \Gls{bs}: $ \gls{not:state}^{\mathrm{b}}_{t} = ( \gls{not:obs}^{\mathrm{b}}_{t}, \ldots , \gls{not:obs}^{\mathrm{b}}_{t-k} , \mathbf{\gls{not:ul-message}}_{t}, \ldots, \mathbf{\gls{not:ul-message}}_{t-k} , \mathbf{\gls{not:dl-message}}_{t}, \ldots, \mathbf{\gls{not:dl-message}}_{t-k} ) $, with $ \mathbf{\gls{not:ul-message}}$ and $\mathbf{\gls{not:dl-message}}$ containing the messages from all the \glspl{ue}.
\end{itemize}

We assume the episode ends when a maximum number of steps \gls{not:episode-duration} is reached.
The reward given at each time step is:
\begin{equation}
\label{eq.:reward}
    r_{t} = \begin{cases}
    +\gls{not:reward-signal}, \text{ if a new \gls{sdu} was received by the \gls{bs} } \\
    -\gls{not:reward-signal}, \text{ } \parbox[t]{.25\textwidth}{ if an \gls{ue} deleted a \gls{sdu} that has not been received by the \gls{bs}} \\
    0, \text{ else,}
\end{cases}
\end{equation}
\noindent where \gls{not:reward-signal} is a positive integer.
This choice of reward is possible by leveraging the \gls{ctde}.
During the centralized training, a centralized reward system can be used to observe the buffers of the \gls{bs} and \glspl{ue} in order to assign the reward.
For wireless systems, centralized training can be achieved in a simulation environment as well as a testbed, i.e. a server farm.
\subsection{Training Algorithm}
\label{sec:solution}

The proposed \gls{rl} solution is based on the \gls{maddpg} algorithm \cite{lowe2017multi}.
This algorithm is well suited to partially observable environments when strong coordination is needed, due to its centralized critic architecture.

In \gls{maddpg}, each agent has an actor network that depends only on its own agent's state in order to learn a decentralized policy $\mu_i$ with parameters $\theta_i$.
Each agent also has a centralized critic network that receives the agent states and actions of all agents in order to learn a joint action value function $Q_i (x,a)$ with parameters $\varphi_i$, where $x = (x_1, x_2, \ldots, x_n)$ is a vector containing all the agents' states and $a = (a_1, a_2, \ldots, a_n)$ contains the actions taken by all of the agents.
The critic networks are only used during the centralized training.

The critic network parameters $\varphi$ are updated by minimizing the loss given by the temporal-difference error
\begin{equation}
\label{eq.:critic}
    L^i \coloneqq \E_{x, a, r, x^{\prime} \thicksim \mathcal{D} } \left[ \left( y^i - Q_i (x, a_1, \ldots, a_n ; \varphi_i) \right) ^2  \right]
\end{equation}
\noindent where $\mathcal{D}$ denotes the experience replay buffer in which the transition tuples $ (x, a, r, x^{\prime})$ are stored, $Q^{\prime}$ and $\mu^{\prime}$ represent the target critic network and the value of the target actor network, with parameters $\theta^{\prime}$ and $\varphi^{\prime}$, respectively, and $y^i$ is the temporal-difference target, given by
\begin{equation}
\label{eq.:td-error}
    y^i \coloneqq r + \left. \gamma Q^{\prime}_i ( x^{\prime}, a^{\prime}_1,  \ldots, a^{\prime}_n; \varphi^{\prime}_i )  \right|_{a^{\prime}_k = \mu^{\prime}_k (x_k)}
\end{equation}
\noindent where $\gamma$ is the discount factor.
The actor network parameters $\theta$ are updated using the sampled policy gradient
\begin{equation}
\label{eq.:actor}
    \nabla_{\theta_i} J = \E_{x, a \thicksim \mathcal{D} } \left[ \nabla_{a_i} Q_i (x,a) \nabla_{\theta_i} \mu_i (x_i) \mid a_i=\mu_i (x_i) \right] .
\end{equation}
The target networks parameters are updated as
\begin{equation}
\label{eq.:target-critic}
    \varphi^{\prime}_i  \leftarrow \tau \varphi_i + (1-\tau) \varphi^{\prime}_i
\end{equation}
\begin{equation}
\label{eq.:target-actor}
    \theta^{\prime}_i  \leftarrow \tau \theta_i + (1-\tau) \theta^{\prime}_i
\end{equation}
\noindent where $\tau \in \left[ 0 , 1 \right]$ is the soft-update parameter.

\paragraph*{Architecure}
The actor and critic networks have the same architecture; a fully connected \gls{mlp} with two hidden layers, of $64$ neurons each.
The activation function of all hidden layers is the \gls{relu}.
In order to improve training of our \gls{maddpg} solution, we make use of parameter sharing \cite{foerster2016learning} for similar network nodes, in this case the \glspl{ue}.
Since \gls{ue} index is not included in the agent's state, any policy that uses the agent's identity is not capable of effectively solving the task due to the parameter sharing, because it would lead to collisions.

Similarly to the original work \cite{lowe2017multi}, we use the Gumbel-softmax \cite{gumbel-softmax} trick to soft-approximate the discrete actions to continuous ones.
The Gumbel-softmax reparameterization also works to balance exploration and exploitation.
The exploration-exploitation trade-off is controlled by the temperature factor $\zeta$.

After training finishes, we have successfully trained a population of $\gls{not:repetitions}$ protocols.
We then select the protocol that performed the best at any point during training across all different protocols, i.e the historically best protocol.
This selection step can be seen as a "survival of the fittest" approach because only one protocol of the population of \gls{not:repetitions} is chosen going forward.

\section{Results}
\label{sec:simulation}
\subsection{Simulation Procedure and Parameters}

The transmission buffer of each user starts empty and the \gls{sdu} arrival probability is \gls{not:p-arrival} for each \gls{ue}.
The system is trained for a fixed number of episodes \gls{not:episodes-train}.
During training, we evaluate the policy on a fixed set of \gls{not:episodes-eval} evaluation episodes with disabled exploration and disabled learning in order to assess the current performance of the communication protocol.
The protocol that performed the best on the evaluation episodes during the whole training procedure is selected and its performance is assessed in \gls{not:episodes-test} episodes with exploration and learning disabled.
This whole procedure represents a single training repetition.
We evaluate a total of \gls{not:repetitions} repetitions, each with a different random seed.
A summary of the main simulation parameters is provided in Table~\ref{tab:sim-params}, while the parameters of the \gls{maddpg} and \gls{ddpg} algorithms are listed in Table~\ref{tab:rl-params}.
\begin{table}[tb]
\centering
\caption{Simulation Parameters}
\label{tab:sim-params}
\begin{tabularx}{0.98\columnwidth}{l c c}
\toprule
\textbf{Parameter}                      & \textbf{Symbol} 	             & \textbf{Value}           \\
\midrule
Number of \glspl{ue}                    & \gls{not:nUE}                 &  $\left[2, 3, 4, 5 \right]$                      \\
Size of transmission buffer             & \gls{not:buffer-size}         &  $20$                      \\
Avg. number of \glspl{sdu} per \gls{ue}       & \gls{not:avg-packets}       &  $\left[2, 4, 6, 8, 10, 12 \right]$    \\
\Gls{sdu} arrival probability           & \gls{not:p-arrival}           &  \makecell{$ \left[ 0.083, 0.16, 0.25, \right. $ \\ $ \left. 0.33, 0.41, 0.5 \right] $}                    \\
Transport block error rate              & \gls{tbler}                   &  $10^{-1}$               \\
\gls{dcm} vocabulary size               & \gls{not:dl-vocabulary-size}  &  $3$                 \\
\gls{ucm} vocabulary size               & \gls{not:ul-vocabulary-size}  &  $2$                 \\
Duration of episode (\glspl{tti})  & \gls{not:episode-duration}    &  $24$                \\
Reward function parameter               & \gls{not:reward-signal}       &  $3$                 \\
Number of training episodes             & \gls{not:episodes-train}      &  $100 \si{k}$        \\
Number of evaluation episodes           & \gls{not:episodes-eval}       &  $500$               \\
Number of test episodes                 & \gls{not:episodes-test}       &  $5000$              \\
Number of randomized repetitions        & \gls{not:repetitions}         &  $8$                \\
\bottomrule
\end{tabularx}
\end{table}

\begin{table}[tb]
\centering
\caption{Training Algorithm Parameters}
\label{tab:rl-params}
\begin{tabularx}{0.98\columnwidth}{l c c}
\toprule
\textbf{Parameter}                      & \textbf{Symbol}       & \textbf{Value} 	            \\
\midrule
Memory length                           & $k$      & 3                             \\
Replay buffer size                      &       & $10^5$                        \\
Batch size                              &       & 1024                          \\
Number of neurons per hidden layer      &       & $\{ 64, 64 \} $      \\
Interval between updating policies      &       & 96                            \\
Optimizer algorithm                     &       & Adam    \\
Learning rate                           & $\alpha$      & $10^{-3}$                     \\
Discount factor                         & $\gamma$     & $0.9$                        \\
Policy regularizing factor              &       & $10^{-3}$                     \\
Gumbel-softmax temperature factor       & $\zeta$       & $1$                           \\
Target networks soft-update factor      & $\tau$      & $10^{-3}$                     \\

\bottomrule
\end{tabularx}
\end{table}

\subsection{Baseline Solutions}
We compare the proposed solution with a contention-free (i.e. \gls{bs}-controlled, scheduled) and a contention-based (i.e. grant-free) baseline.

\begin{figure*}[!tb]
    \centering
    \begin{subfigure}[t]{0.45\textwidth}
        \centering
        \includegraphics[width=\textwidth]{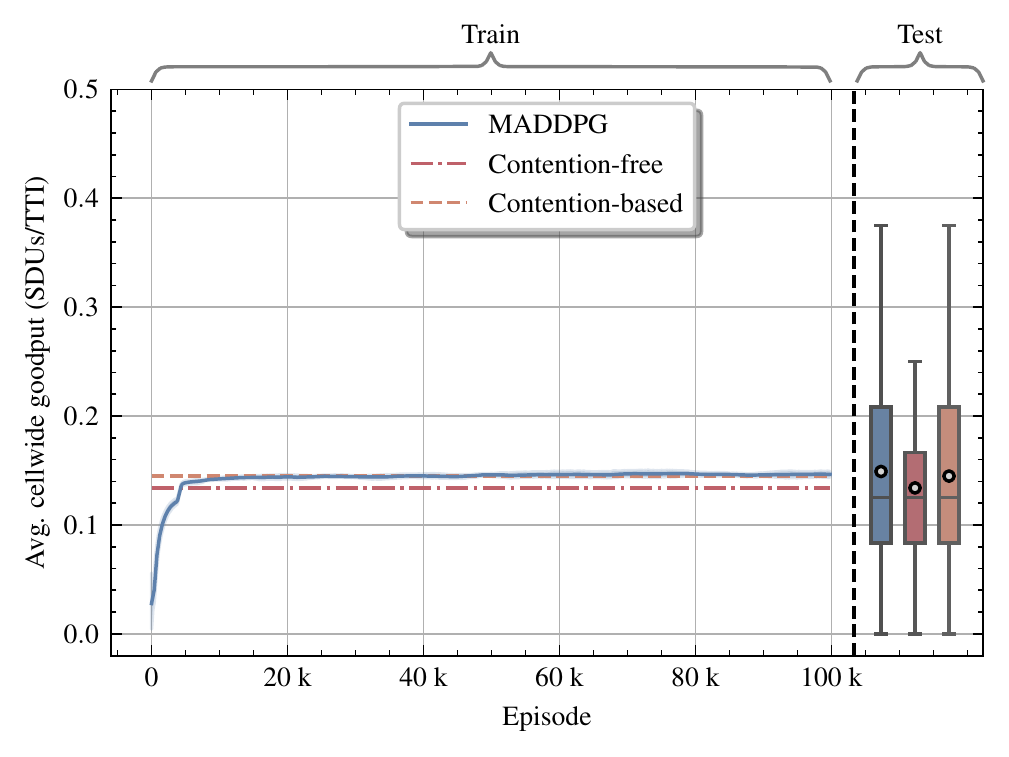}
        \caption{Avg. Number of SDUs per \gls{ue}: $\gls{not:avg-packets}=2$. Arrival rate: $\gls{not:p-arrival}=0.83$.}
        \label{fig:learning-1}
        \end{subfigure}%
        ~
        \begin{subfigure}[t]{0.45\textwidth}
            \centering
            \includegraphics[width=\textwidth]{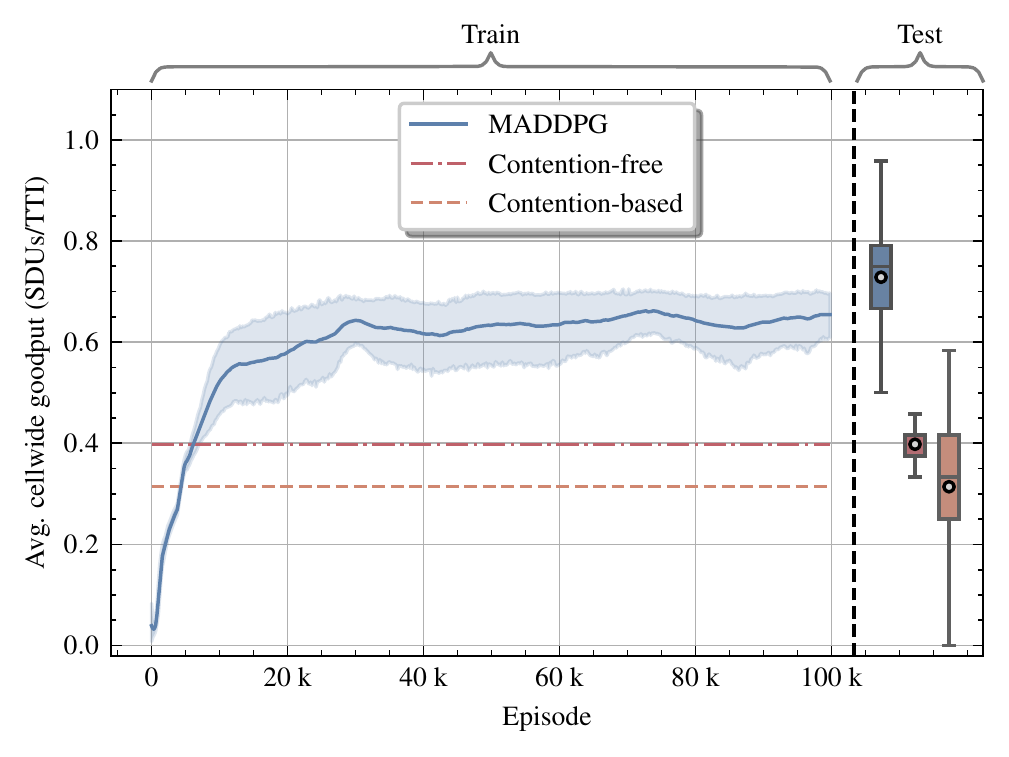}
            \caption{Avg. Number of SDUs per \gls{ue}: $\gls{not:avg-packets}=12$; Arrival rate: $\gls{not:p-arrival}=0.5$.}
            \label{fig:learning-2}
        \end{subfigure}
        \caption{Goodput comparison during the training procedure. Number of \glspl{ue}: $\gls{not:nUE} = 2$ ; \gls{tbler} $= 10^{-1}$. }
        \label{fig:train-results-goodput}
    \end{figure*}

In the contention-free protocol, the \gls{ue} sends a \gls{sr} if its transmission buffer is not empty and it only transmits if it has received a \gls{sg}.
Similarly, it only deletes a \gls{tb} from the transmission buffer after the reception of an \gls{ack}.
At each time step, the \gls{bs} receives zero or more \glspl{sr}.
It then chooses one of the requesters at random to transmit in the next time-step, sending a \gls{sg} to the selected \gls{ue}.
However, if the \gls{ue} had made a successful data transmission simultaneously with an \gls{sr}, the \gls{bs} will send an \gls{ack} to this \gls{ue} and its \gls{sr} is ignored.

In the contention-based protocol, each \gls{ue} transmits with probability $p_{t}$ if its transmission buffer is not empty.
Similarly to the contention-free baseline, the \gls{ue} only deletes a \gls{tb} after the reception of an \gls{ack}.
At each time step, the \gls{bs} sends an \gls{ack} to a \gls{ue} if it received a \gls{tb} from the \gls{ue}.
For each experiment, the transmission probability chosen is the one that performs better in terms of goodput.

\subsection{Results}

    \begin{figure*}[!tb]
        \centering
        \begin{subfigure}[t]{0.45\textwidth}
            \centering
            \includegraphics[width=\textwidth]{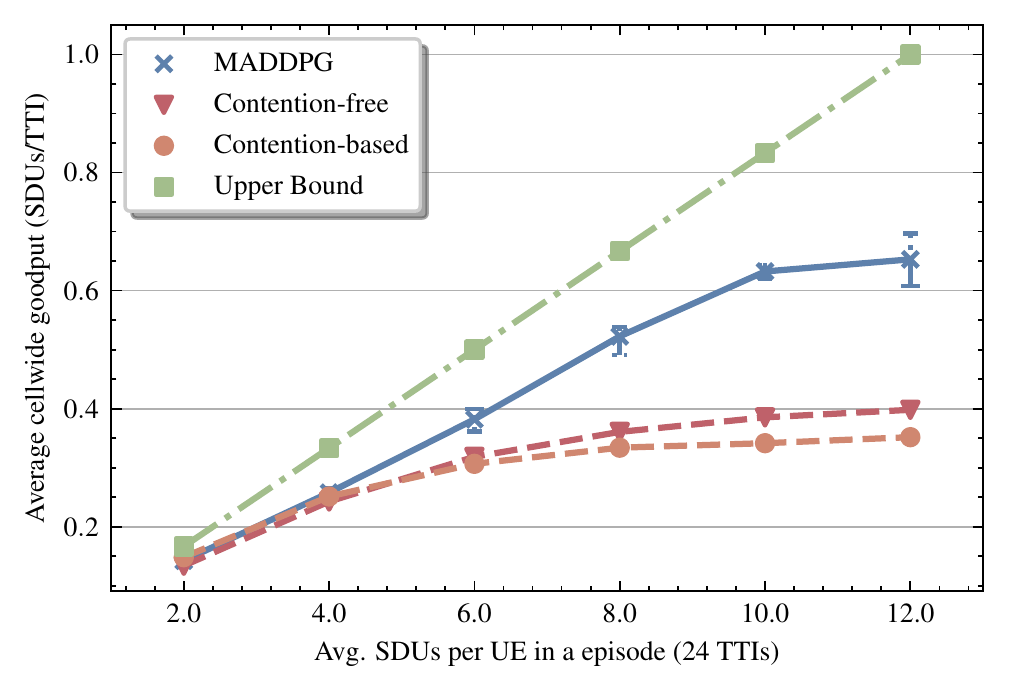}
            \caption{Goodput when increasing $\gls{not:p-arrival}$.}
            \label{fig:goodput-arrival}
        \end{subfigure}
        ~
        \begin{subfigure}[t]{0.45\textwidth}
            \centering
            \includegraphics[width=\textwidth]{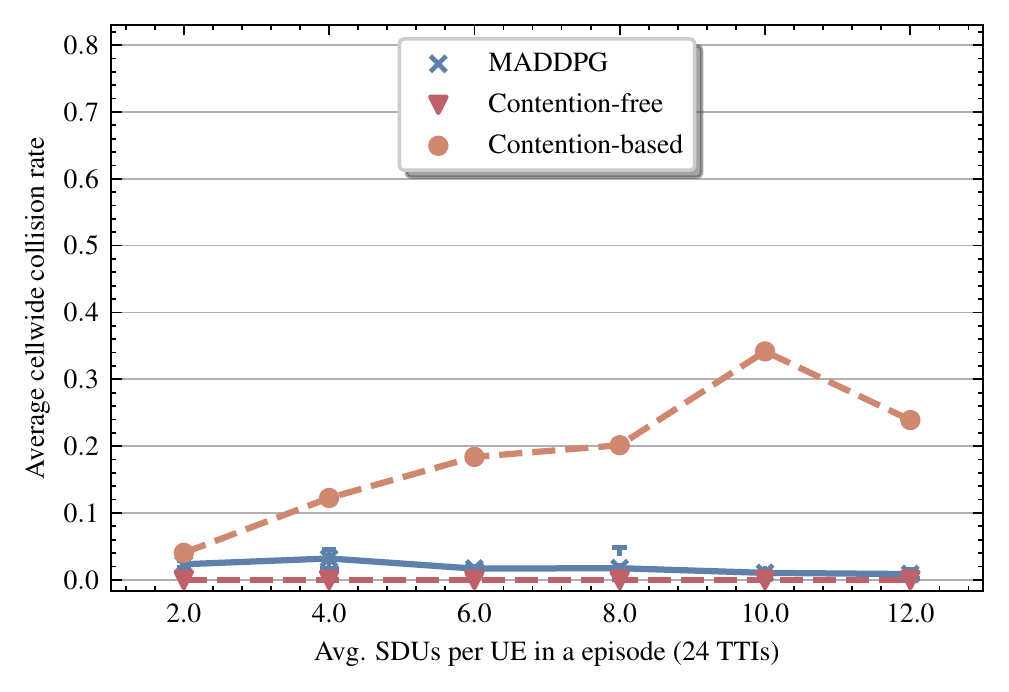}
            \caption{Collision rate when increasing $\gls{not:p-arrival}$.}
            \label{fig:collision-arrival}
            \end{subfigure}%
        \caption{Scaling the arrival rate while mantaining the number of \glspl{ue} fixed: \gls{not:nUE} = 2 \glspl{ue}.}
        \label{fig:scale-arrival}
        \end{figure*}

    \begin{figure*}[!tb]
        \centering
        \begin{subfigure}[t]{0.45\textwidth}
            \centering
            \includegraphics[width=\textwidth]{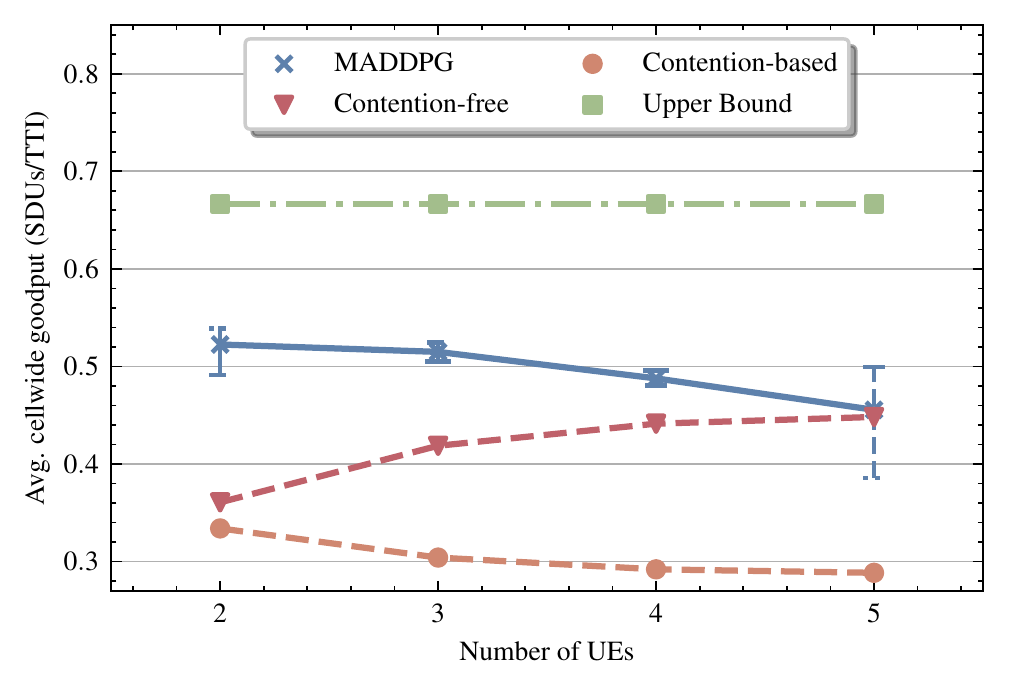}
            \caption{Goodput when increasing the number of \glspl{ue}.}
            \label{fig:goodput-ue}
        \end{subfigure}
        ~
        \begin{subfigure}[t]{0.45\textwidth}
            \centering
            \includegraphics[width=\textwidth]{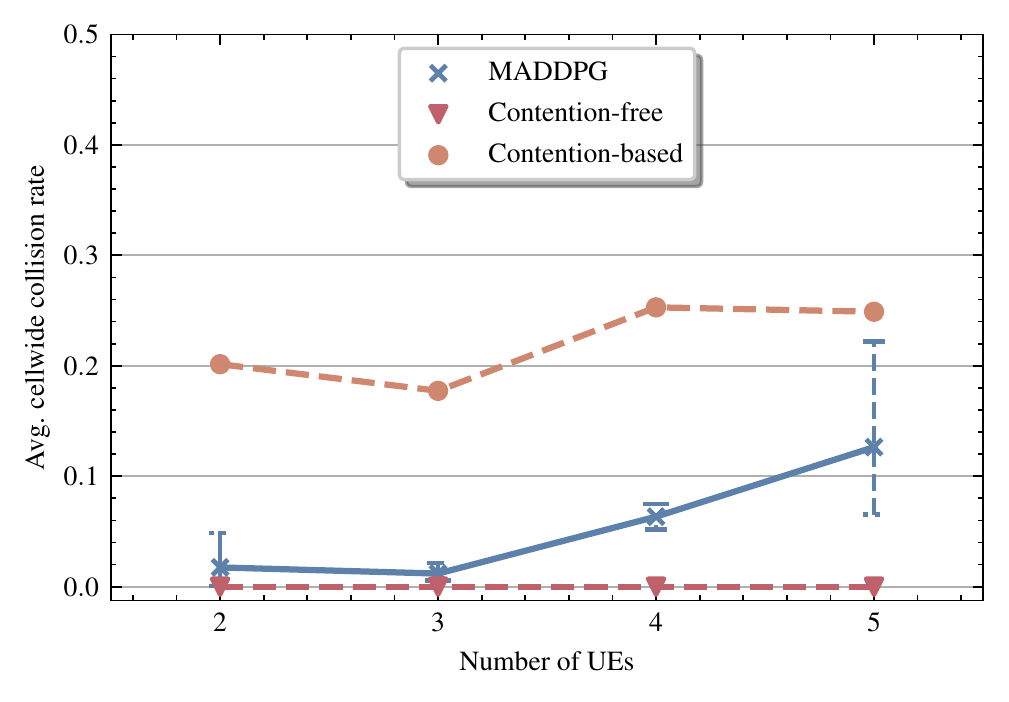}
            \caption{Collision rate per number of \glspl{ue}.}
            \label{fig:collision-ue}
            \end{subfigure}%
        \caption{Scaling the number of \glspl{ue} while maintaining the average number of \glspl{sdu} in the cell per episode fixed: $\gls{not:nUE} \gls{not:avg-packets} = 16 $.}
        \label{fig:scale-ue}
        \end{figure*}

\subsubsection{Learning Performance}
We first analyze the performance over the training procedure, comparing the proposed solution with the baselines in Fig.~\ref{fig:train-results-goodput}.
The solid lines in Figs.~\ref{fig:learning-1} and~\ref{fig:learning-2} show the average performance in the evaluation episodes during the training and the shaded areas represent the $95 \%$ \gls{ci}.
After assessing the performance on the last \gls{not:episodes-eval} evaluation episodes, we select the best performing repetitions for each solution in terms of average goodput to compare using boxplots of the test episodes.

The main conclusions we can draw from Fig.~\ref{fig:train-results-goodput} are:
\begin{itemize}
    \item In the lower arrival rate showed in Fig.~\ref{fig:learning-1}, the proposed solution seems to learn a protocol that performs like a contention-based one. This conclusion is supported by the similar box plots on the test episodes.
    \item In higher arrival rates showed in Fig.~\ref{fig:learning-2}, the proposed solution drastically outperforms both baselines, which indicates it learns a completely different protocol.
    \item The contention-free baseline shows a better performance on low arrival rates, but when the arrival probability increases the contention-free baseline outperforms it.
\end{itemize}

\subsubsection{Scalability}

In this set of results, we analyze the scaling capabilities of the proposed solution across two dimensions, the number of \glspl{ue} and the \gls{sdu} arrival rate.
The performance is evaluated on \gls{not:episodes-test} test episodes by comparing the average goodput and collision rate achieved when changing across one dimension while the other is fixed.
For the \gls{maddpg} solution, we also show the $95\%$ \gls{ci} across randomized repetitions.
The upper bound shows the maximum average goodput when all the \glspl{sdu} are received.

The proposed framework is capable of producing protocols that outperform both baselines in terms of goodput when the arrival rate increases while maintaining a low collision rate, as shown in Fig.~\ref{fig:scale-arrival}.
Also, the \gls{ci} increases when the arrival rate increasing, indicating that in more difficult conditions there's a bigger variability in the emerged protocol.

Scalability to growing numbers of \glspl{ue} is proving challenging, as shown in Fig.~\ref{fig:scale-ue}.
The proposed framework consistently outperforms both solutions for up to four \glspl{ue} and has similar performance to the contention-free solution on average for five \glspl{ue}, but it is unable to scale as well as it does when scaling with traffic.
Although the proposed solution achieves a lower collision rate than the contention-based solution, it seems unable to effectively deal with more \glspl{ue} while avoiding collisions, which can explain why the cellwide goodput drops when increasing the number of \glspl{ue}.

\section{Conclusions and Perspectives}
\label{sec:conclusion}

We have applied a framework to emerge a \gls{mac} protocol and have demonstrated through simulations that cooperative \gls{marl} augmented with communication provides an original approach to emerge a protocol by jointly learning the channel access policy and its signaling.
The results indicate the capabilities of the \gls{marl} framework to produce protocols that outperform the baselines.
In addition, the results illustrate the capabilities of the framework to adapt to different arrival rates and to different number of \glspl{ue}.

In our future works we will propose extensions to deal with even more \glspl{ue}.
We will also propose comparisons of different \gls{marl} algorithms, and we will evaluate accurately the impact of the vocabulary sizes used in the control channels.
Additionally, interpretability will be investigated to better understand the key for improvements used by our \gls{rl} based algorithms.

\section*{Acknowledgment}
\label{sec:ack}
\addcontentsline{toc}{section}{Acknowledgment}
The work of Mateus P. Mota is funded by Marie Sklodowska-Curie actions (MSCA-ITN-ETN 813999 WINDMILL).

\bibliographystyle{IEEEtran}
\bibliography{IEEEabrv, ref.bib}

\begin{thebibliography}{10}
\providecommand{\url}[1]{#1}
\csname url@samestyle\endcsname
\providecommand{\newblock}{\relax}
\providecommand{\bibinfo}[2]{#2}
\providecommand{\BIBentrySTDinterwordspacing}{\spaceskip=0pt\relax}
\providecommand{\BIBentryALTinterwordstretchfactor}{4}
\providecommand{\BIBentryALTinterwordspacing}{\spaceskip=\fontdimen2\font plus
\BIBentryALTinterwordstretchfactor\fontdimen3\font minus
  \fontdimen4\font\relax}
\providecommand{\BIBforeignlanguage}[2]{{%
\expandafter\ifx\csname l@#1\endcsname\relax
\typeout{** WARNING: IEEEtran.bst: No hyphenation pattern has been}%
\typeout{** loaded for the language `#1'. Using the pattern for}%
\typeout{** the default language instead.}%
\else
\language=\csname l@#1\endcsname
\fi
#2}}
\providecommand{\BIBdecl}{\relax}
\BIBdecl

\bibitem{6g_roadmap}
K.~B. Letaief, W.~Chen, Y.~Shi, J.~Zhang, and Y.-J.~A. Zhang, ``The roadmap to
  {6G}: {AI} empowered wireless networks,'' \emph{IEEE Communications
  Magazine}, vol.~57, no.~8, pp. 84--90, 2019.

\bibitem{foerster2016learning}
J.~N. Foerster, Y.~M. Assael, N.~de~Freitas, and S.~Whiteson, ``Learning to
  communicate with {D}eep multi-agent reinforcement learning,'' in
  \emph{Proceedings of the 30th International Conference on Neural Information
  Processing Systems}, 2016, pp. 2145--2153.

\bibitem{sukhbaatar2016learning}
S.~Sukhbaatar, A.~Szlam, and R.~Fergus, ``Learning multiagent communication
  with backpropagation,'' in \emph{Proceedings of the 30th International
  Conference on Neural Information Processing Systems}, 2016, pp. 2252--2260.

\bibitem{dutta2021Synthesis}
H.~Dutta and S.~Biswas, ``Towards multi-agent reinforcement learning for
  wireless network protocol synthesis,'' in \emph{2021 International Conference
  on COMmunication Systems NETworkS (COMSNETS)}, 2021, pp. 614--622.

\bibitem{guo2022Access}
Z.~Guo, Z.~Chen, P.~Liu, J.~Luo, X.~Yang, and X.~Sun, ``Multi-agent
  reinforcement learning based distributed channel access for next generation
  wireless networks,'' \emph{IEEE Journal on Selected Areas in Communications},
  pp. 1--1, 2022.

\bibitem{gomes2020automatic}
A.~Gomes, D.~F. Macedo, and L.~F. Vieira, ``Automatic mac protocol selection in
  wireless networks based on reinforcement learning,'' \emph{Computer
  Communications}, vol. 149, pp. 312--323, 2020.

\bibitem{pasandi2021}
H.~B. Pasandi and T.~Nadeem, ``Towards a learning-based framework for
  self-driving design of networking protocols,'' \emph{IEEE Access}, vol.~9,
  pp. 34\,829--34\,844, 2021.

\bibitem{alvaro2021}
A.~Valcarce and J.~Hoydis, ``Towards joint learning of optimal {MAC} signaling
  and wireless channel access,'' \emph{IEEE Transactions on Cognitive
  Communications and Networking}, pp. 1--1, 2021.

\bibitem{mota2021emergence}
M.~P. Mota, A.~Valcarce, J.-M. Gorce, and J.~Hoydis, ``The emergence of
  wireless mac protocols with multi-agent reinforcement learning,'' in
  \emph{2021 IEEE Globecom Workshops (GC Wkshps)}.\hskip 1em plus 0.5em minus
  0.4em\relax IEEE, 2021, pp. 1--6.

\bibitem{oliehoek2008optimal}
F.~A. Oliehoek, M.~T. Spaan, and N.~Vlassis, ``Optimal and approximate q-value
  functions for decentralized pomdps,'' \emph{Journal of Artificial
  Intelligence Research}, vol.~32, pp. 289--353, 2008.

\bibitem{lowe2017multi}
R.~Lowe, Y.~I. Wu, A.~Tamar, J.~Harb, O.~P. Abbeel, and I.~Mordatch,
  ``Multi-agent actor-critic for mixed cooperative-competitive environments,''
  in \emph{Advances in neural information processing systems}, 2017, pp.
  6379--6390.

\bibitem{gumbel-softmax}
\BIBentryALTinterwordspacing
E.~Jang, S.~Gu, and B.~Poole, ``Categorical reparameterization with
  {Gumbel-Softmax},'' in \emph{5th International Conference on Learning
  Representations, {ICLR} 2017, Toulon, France, April 24-26, 2017, Conference
  Track Proceedings}.\hskip 1em plus 0.5em minus 0.4em\relax OpenReview.net,
  2017. [Online]. Available: \url{https://openreview.net/forum?id=rkE3y85ee}
\BIBentrySTDinterwordspacing

\end{thebibliography}

\end{document}